\documentclass{aa520}
\usepackage{graphicx}
\usepackage{txfonts}
\newcommand{\ea}{\hbox{et al.}}

\begin{document}
   \title{Are there MACHOs in the Milky Way halo?}


   \author{A. M. Green
          \inst{1,} \inst{2}
          \and
          K. Jedamzik\inst{3,} \inst{4}}

   \offprints{amg@physto.se}

   \institute{Department of Physics, Stockholm University, S--106 91 
            Stockholm, Sweden\thanks{Present address}
              \email{amg@physto.se}
         \and Astronomy Unit, School of Mathematical Sciences, Queen Mary
University of London, Mile End Road, London, E1 4NS, UK
        \and Laboratoire de Physique Math\'ematique et Th\'eorique, 
Universit\'e de Montpellier II, 34095 Montpellier Cedex 5, France\thanks{Present address}
           \email{jedamzik@lpm.univ-montp2.fr}
      \and 
      Max-Planck-Institut f\"ur Astrophysik,
85740 Garching, Germany       }

   \date{Received ; accepted }

   \abstract{Microlensing searches aim to detect compact halo dark
matter via its gravitational lensing effect on stars within the Large
Magellanic Cloud. These searches have led to the claim that 
roughly one fifth of the galactic halo dark matter may be in the form
of compact, solar-mass objects. We analyze this hypothesis by
considering the goodness-of-fit of the best-fit halo dark matter
solutions to the observational data.  We show that the distribution of
the durations of the observed microlensing events is significantly
narrower than that expected to result from a standard halo lens
population at 90 to 95$\%$ confidence, casting doubt on the lenses
constituting halo dark matter.  This conclusion may possibly be
avoided if (i) the Milky Way halo is sufficiently nonstandard or (ii)
a large fraction of the events are due to non-halo populations with
event durations coincidentally close to those of the putative halo
population or (iii) individual event durations have been seriously
underestimated due to blending.

   }

   \maketitle
%

\section{Introduction}
Massive compact halo objects (MACHOs) with mass in the range $10^{-8}
M_{\odot}$ to $10^{3} M_{\odot}$ can be detected via the temporary
amplification of background stars which occurs, due to gravitational
microlensing, when the MACHO passes close to the line of sight to a
background star (Paczy\'{n}ski~\cite{pac}). In the early 1990s several
collaborations began monitoring millions of stars in the Large and
Small Magellanic Clouds (LMC and
SMC)~\footnote{http://wwwmacho.mcmaster.ca/, http://eros.in2p3.fr/},
in order to search for Milky Way (MW) compact halo dark matter, and
indeed a number of candidate microlensing events have been observed
(Alcock et al.~\cite{5.7years}; Lasserre et al.~\cite{eros}; Milsztajn
\& Lasserre~\cite{millas}).  The interpretation of these microlensing
events is a matter of much debate. Though the lenses may easily reside
within the MW halo and constitute halo dark matter, it is also
possible that the contribution to the lensing rate due to other, so
far unknown non-halo populations of objects has been underestimated
(see e.g.  Bennett~\cite{nomacho}; Zhao~\cite{nomacho2}).  In
particular the lensing rate due to stars, or MACHOs, within the LMC
itself (Wu~\cite{wu}; Auborg et al.~\cite{sl1};
Salati et al.~\cite{sl2}; Evans \& Kerins~\cite{sl3}; Gyuk, Dalal \&
Griest~\cite{gg}), or a very thick Milky Way disk
(Gyuk \& Gates~\cite{gg01}) may be significant.

The EROS collaboration have five candidate events, resulting from
three years of observation of 25 million stars in the LMC (Milsztajn \&
Lasserre~\cite{millas}). Due to the small number of events, and poor
fit to microlensing of several of these events, they use these events
to place constraints on the halo fraction as a function of lens mass,
assuming a standard Maxwellian MW halo. In contrast, the MACHO
collaboration claim that roughly 20$\%$ of the halo is in compact
objects with $M \sim 0.5 M_{\odot}$, using their 13/17 candidates
events which result from the 5.7 years of observation of 10.7 million
stars (Alcock et al.~\cite{5.7years}). While not in direct conflict
with the EROS exclusion limits these MACHO best fit parameters lie
just outside the EROS exclusion limits. These differences could be due
to EROS covering a larger solid angle than MACHO, such that the
lensing rate due to lenses in the LMC itself should be smaller, as
well as the use of less crowded fields by EROS as compared to MACHO,
simplifying the estimation of the event durations. See Milsztajn
(\cite{mil}) for an extended discussion of the current observational
situation.

In this paper we revisit the analysis of the MACHO collaborations
candidate microlensing events (Alcock et al.~\cite{5.7years}). We
contemplate whether the advocated interpretation of the microlensing
observations in terms of a MACHO halo dark matter component with
Maxwellian velocity distribution within a standard MW halo and with
best-fit typical mass $0.6\,\, M_{\odot}$ and halo dark matter
fraction $ f \approx 20\% \,\, $ actually provides, in absolute terms,
a good fit to the data.

\section{Candidate events}
The MACHO collaboration apply two sets of selection criteria to their
data. Criteria A is the most restrictive, only accepting events with a
single highly significant bump in their lightcurve as expected for
simple, single lens, events, and results in 13 events. Criteria B is
looser with 17 events passing the selection criteria, being designed
to also accept low signal to noise events and exotic events, such as
those where the lens is binary and the lightcurve has distinctive
caustic features.  Marginal events which are suspected of being
supernovae occurring in host galaxies behind the LMC are rejected from
criteria A but kept in criteria B.  Further study of event number
22~\footnote{We use the MACHO collaboration's event numbering
nomenclature (Alcock et al.~\cite{5.7years}) throughout.} has found
that the source is extended and has emission lines uncharacteristic of
a stellar object and it is hence very unlikely a microlensing event
(Alcock et al.~\cite{long}). This extraordinarily long event will
henceforth be excluded from the sample, leaving 16 candidate events
for criteria B.

Known stellar populations within the MW disk and bulge, as well as
within the LMC disk, are expected to account for approximately one
quarter of the observed events (Alcock et
al.~\cite{5.7years}). Follow-up observations of fields containing each
of the source stars have been carried out using the Hubble Space
Telescope Wide Field Planetary Camera 2 (Alcock et
al.~\cite{nature}). In the case of event number 5 (which passes both
sets of selection criteria) a faint red object has been detected close
to the source star. A chance super-position is extremely unlikely and
a combined analysis of the HST and microlensing data finds that this
object is most likely an M dwarf located in the MW disk.

It is possible to obtain more information about the properties of a
given lens if either the lens or source is within a binary, as the
microlensing lightcurve then exhibits additional features. The lens
responsible for event 9 is a binary, producing distinctive caustic
features in the lightcurve from which the source crossing radius and
hence the lens projected velocity could be estimated (Bennett et
al.~\cite{event9}; Alcock et al.~\cite{binary}). The low projected
velocity found suggests that either the lens resides in the LMC disk
or the source star is also binary. Accurate photometry of event number
14 (which passes both sets of selection criteria) revealed a periodic
modulation in the lightcurve. Fits to a rotating binary source have
been performed which indicate that it is most likely that the lens
resides in the LMC disc, though direct spectral observations of the
source are required to confirm this (Alcock et al.~\cite{event14}).

While there is clear evidence that event number 22 is not due to
microlensing and that event 5 is due to a non MW halo lens, the
locations of the lenses responsible for events 9 and 14 are not
unambiguously known. Furthermore no conclusions on the locations of
the entire lens population can be drawn from the locations of the
lenses responsible for events 9 and 14, due to the biases involved in
observing exotic events (Alcock et al.~\cite{event14};
Honma~\cite{honma}).

 A further complication is the estimation of the event durations. In
crowded stellar fields it is not possible to resolve individual stars
and some fraction of the baseline flux may come from unlensed
stars. This phenomena, known as blending, results in an underestimate
of the event timescale. The MACHO collaboration take this into account
by carrying out sophisticated Monte Carlo simulations to find the
average factor by which the real timescale is underestimated, and then
correcting each of the fitted durations by this factor (Alcock et
al.~\cite{5.7years}; Alcock et al.~\cite{long}).

\section{Likelihood analysis}
The MACHO collaboration defines the likelihood of a given model as the
product of the Poisson probability of observing $N_{{\rm obs}}$ events
when expecting $N_{{\rm exp}}$ events and the probabilities of finding
the observed durations $\hat{t}_{{\rm j}}$ (where $j=1,....,N_{{\rm
obs}}$) from the theoretical duration distribution (Alcock et
al.~\cite{likeref}; Alcock et al.~\cite{likeref2}):
\begin{equation}
\label{likedef}
{\mathcal L} = \exp{\left(-N_{{\rm exp}}\right)} \Pi^{N_{{\rm obs}}}_{j=1}
\mu_{{\rm j}} \, , 
\end{equation}
where the expected number of events is given by
\begin{equation}
\label{nexp}
N_{{\rm exp}} = E \int_{0}^{\infty} \frac{{\rm d} \Gamma}{{\rm d} t}
           \,  \epsilon(t) \, {\rm d} t \, ,
\end{equation}
and $\mu_{{\rm j}}$ by
\begin{equation}
\mu_{{\rm j}} = E \, \epsilon(t_{{\rm j}}) \, \frac{{\rm d} \Gamma
(t_{{\rm j}})}{{\rm d} t} \, .
\end{equation}
Here $E=6.12 \times 10^{7}$ star-years is the exposure, $\epsilon(t)$
is the detection efficiency, $t$ is the estimated event duration
statistically corrected for blending and ${\rm d} \Gamma /{\rm d} t$
is the differential event rate. The likelihood approach can be used to
compare how well different sets of parameter values within a {\em
given} theoretical model fit the data. In the present context these
parameters are the MACHO mass and halo fraction and the model is a
standard halo population of compact dark matter.  However, the
likelihood approach does not tell one, in absolute terms, {\em how
good} a fit really is. In particular, it is possible that though
"best-fit" model parameters have been found, the model provides a
generally bad description of the observational data.  In order to
illustrate this we compare ${\rm d} \Gamma /{\rm d} t$ of the best fit
standard halo model with delta MACHO mass function (S-DMF) with that
of a best fit (i.e. maximum likelihood) Gaussian differential event
rate with variable amplitude, center and width.  We note that the
Gaussian has no physical motivation and is intended to simply provide
some arbitrary reference to the S-DMF model.~\footnote{Note that the
Gaussian model has one more fitting parameter than the S-DMF model.}

The best fit Gaussian and S-DMF differential event rates which arise
in a likelihood analysis assuming that all 13 events found using
criteria A are due to MW halo lenses (i.e. neglecting the event rate
due to background events) are plotted in Fig. 1 multiplied by the
detection efficiency. The Gaussian clearly gives a better fit to the
observed event distribution than the S-DMF model, and its likelihood
is greater by a factor of 32.  We have also carried out this
comparison for all possible 10 event sub-sets (i.e. subtracted every
possible combination of 3 background events) and in every case the
Gaussian has a larger maximum likelihood (by a factor ranging from 5
to 4500).

\begin{figure}
\centering
\includegraphics[width=8cm]{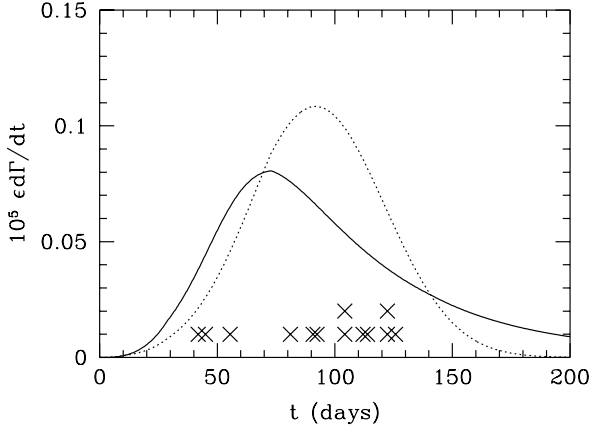}
\caption{The best fit delta function plus standard halo
model (solid line) and Gaussian (dotted) differential event rates for
the 13 events resulting from criteria A. Crosses indicate the observed
event durations. An analogous plot for the 16 events resulting from
criteria B appears very similar.}
\end{figure}

\section{Event duration distribution}

Fig. 1 illustrates the reason why the artificial Gaussian event
duration distribution provides a better fit to the data than the event
duration distribution produced by a MACHO population within a standard
halo model; namely, whereas the S-DMF predicts the occurrence of long
duration events none are observed. To quantify this discrepancy we
have performed a Monte Carlo simulation which compares the observed
width of the distribution
\begin{equation}
\label{tau}
Width = {{{\rm Max}_{\{j=1, N^*_{obs}\}} (t_j) - {\rm Min}_{\{j=1, N^*_{obs}\}} 
(t_j)}\over {\rm Avg}_{\{j=1, N^*_{obs}\}} (t_j)}\, ,
\end{equation}
to the $Width$ obtained by statistically drawing a number
$N^{halo}_{obs}$ of events from a standard halo event duration
distribution (multiplied by the relevant detection efficiency and
taking into account the transverse velocity of the line-of-sight). We
note that the above statistic is designed to be independent of the
best-fit MACHO halo fraction and mass~\footnote{There remains a
residual dependence of the $Width$ on the typical MACHO mass, however,
due to the duration dependence of the detection efficiencies.}  and is
thus a parameterization of the observations which is essentially
orthogonal to the best-fit parameters determined by the MACHO
collaboration~\footnote{This approach is similar to that of de Rujula
et al. (\cite{der}) who showed that the duration dispersion could in
principle be used to separate background events.}.

In our Monte Carlo analysis we find the fraction of generated event
samples that yield a $Width$ which is smaller than that observed. To
account for background (i.e. non-halo) lenses we employ the estimates
for the expected number of background events $<N_{BG}>$ found in
Alcock et al. (\cite{5.7years}). We do not distinguish between the
different background populations, but simply subtract a number
$N^*_{BG}$ from $N^*_{obs}$ to obtain $N^{halo}_{obs}$, the number of
events which are expected to be due to MACHOs and are to be generated
from the theoretical event duration distribution. We do this using two
different methods: subtracting a fixed number $N^*_{BG} = <N_{BG}>^*$,
and subtracting $N^*_{BG}$ as drawn from a Poisson distribution with
average $<N_{BG}>^*$.~\footnote{In some of our samples $<N_{BG}>$ as
given in Alcock et al. (\cite{5.7years}) is adjusted to $<N_{BG}>^*$
to account for events which have been excluded from the sample by
hand, cf. Table 1.}  We note here that while the background events
could potentially account for the longest or shortest events in the
sample, we nevertheless do not change the observed $Width$ on
subtracting background events, such that our estimated probabilities,
in a conservative spirit, should generally reflect an overestimate.

\begin{table}
\centering
\caption[]{Probability of various subsamples of the observed candidate
microlensing events from the 5.7 year candidate events being drawn
from an event duration distribution function (i.e. ${\rm d} \Gamma
/{\rm d} t$) as predicted by the standard MW halo model. Here the
columns indicate the number of candidate events ($N_{obs}=13$ for
efficiency A, and $N_{obs}=17$ for efficiency B), whether a delta-
(DMF) or Gaussian- (GMF) MACHO mass function has been assumed, the
numbers of the events which have been excluded from the sample by hand
(correspondingly reducing the number of observed events from $N_{obs}$
to $N^*_{obs}$), the assumed average number $<N_{BG}>^*$ of events due
to non-halo background (BG) lens populations, as well as the
aforementioned probabilities, computed either by subtracting a fixed
number $<N_{BG}>^*$ from $N^*_{obs}$ to obtain the number of halo
events $N^{halo}_{obs}$ to be drawn from the theoretical event
duration distribution, or by finding the number $N_{BG}^*$ of events
to be subtracted from $N^*_{obs}$, from a Poisson distribution with
average $<N_{BG}>^*$.  In the case of non-integer $<N_{BG}>^*$, this
number is rounded up to the next integer, when fixed number background
subtraction is performed. Note that the exclusion of event number 22
from the sample by hand is not assumed to reduce $<N_{BG}>$ from 3.9
to 2.9, as it is a supernova rather than microlensing due to a
background lensing population.}
\begin{tabular}{cccccc}
\multicolumn{6}{c}{Table 1}\\
\hline \hline
Model/ & mass & events & \# BG & Prob. & Prob. \\
Sample & funct. & excluded & events & (fixed) & (Poisson) \\
\hline
S/13 & DMF & none & 3.0 & 6.9\% & 8.9\%  \\
S/13 & GMF & none & 3.0 & 3.5\% & 5.2\%  \\
S/13 & DMF & 5 & 2.0 & 6.9\% & 9.1\%  \\
S/13 & GMF & 5 & 2.0 & 3.5\% & 5.2\%  \\
S/13 & DMF & 5, 14 & 1.0 & 8.1\% & 8.7\%  \\
S/13 & GMF & 5, 14 & 1.0 & 4.2\% & 4.7\%  \\
S/17 & DMF & 22 & 3.9 & 6.4\% & 7.7\%     \\
S/17 & GMF & 22 & 3.9 & 3.1\% & 4.2\%     \\
S/17 & DMF & 5, 9, 14, 22 & 0.9 & 4.7\% & 4.8\%     \\
S/17 & GMF & 5, 9, 14, 22 & 0.9 & 2.2\% & 2.3\%     \\
\hline
\end{tabular}

\end{table}

The results of this procedure, in particular the fraction of
simulations which have smaller $Width$ than observed, are shown in
Table 1 for a variety of different subsamples. This fraction is
generally between 5\% and 10\%, such that the underlying S-DMF model
serving as an explanation for the observations is ruled out at the
90 to 95\% confidence level. Note that this conclusion does not vary
strongly if the events which are suspected of being due to background
populations are excluded by hand. This would not be the case if event
number 22, which has a duration of 297.8 days, was a genuine
microlensing event.

The discrepancy between the observations and the theoretical
expectations even worsens when the assumption of a delta function MACHO
mass function (DMF) is dropped. To illustrate the effects of a MACHO
mass function with finite width we take the number of MACHOs with mass
within a logarithmic interval to be
\begin{equation}
{{\rm d}\, P\over {\rm d}\, {\rm ln}(M/M^*)} = {1\over\sqrt{2\pi}}{1\over\sigma}
\, {\rm exp}\biggl\{-{1\over 2}\biggl({{\rm ln}\bigl({M\over M^*}
\bigr)\over\sigma}\biggr)^2\biggr\}\, ,
\end{equation}
with
\begin{equation}
\sigma = \sqrt{{\rm ln\, 2}/2}\, ,
\end{equation}
such that MACHOs with either twice or one-half the average mass $<M> =
{\rm exp}\, (\sigma^2/2)\, M^*$ (which is assumed to take the best-fit
value of the DMF) are a factor of two less common. These models are
indicated in the table by GMF.

It may be noted that the probabilities are generally larger when
background event subtraction is performed by employing Poisson
statistics, rather than by subtracting a fixed number of events. This
is due to the probabilities in the Poisson case being dominated by
samples which have $N^*_{BG}$ larger than $<N_{BG}>^*$, 
and have correspondingly smaller $N_{obs}^{halo}$, making a 
comparison of theoretical expectation and observed distribution 
more favorable. Nevertheless, such reasoning makes the implicit and
nontrivial assumption that all background events have event durations 
falling within the range of those of a putative MACHO population. 

One may ask, making this assumption about coincidence in event
durations between different populations, by how much one has to
increase the number of background events, in order to achieve a
"decent" probability (say 20\%) of the observed distribution indeed
being drawn from the distribution expected for the standard halo
model. Fig. 2 shows the probability for different samples, for the
delta mass function (DMF) and that given by Eq. 5 (GMF), indicating
that an increase of $N^*_{BG}$ by about a factor of two would be
necessary in order for this to be the case.  This suggests, on
statistical grounds, that the contribution of the background lens
populations to the observed events has been underestimated, and
consequently the MACHO halo mass fraction overestimated. Nevertheless,
due to the small number of observed events, a significant contribution
to the observed events by a population of MACHOs within a standard MW
halo may not be ruled out.  This is interesting in light of the recent
detection of a population of old very cool white dwarfs which could
comprise a few per-cent (Oppenheimer et al.~\cite{wd}) of the MW dark
halo, see Richer (\cite{wd1}) for a critical discussion of these
observations.

\begin{figure}
 \centering
\includegraphics[width=8cm]{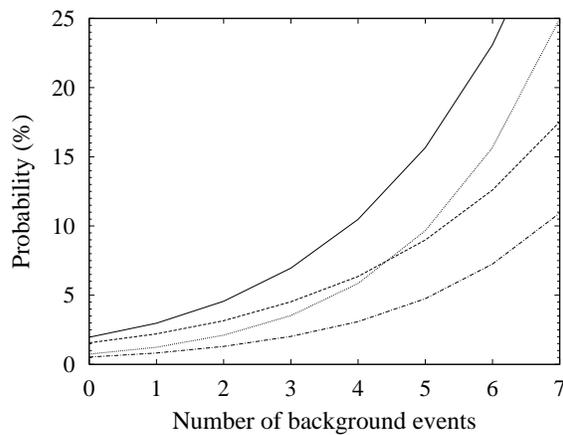}
\caption{Probability of obtaining an event duration distribution for a
standard halo model as narrow as that observed in the 5.7 year sample
as a function of the number of events assumed to be due to
background population(s). Here the background events are implicitly assumed to
neither contribute the longest nor shortest event, but to fall, by pure
coincidence, within the range of the event durations of the putative MACHO
population. Background subtraction is performed by subtracting a fixed (non-Poissonian)
number of events from the number of observed candidate events. The
lines show results for model/sample S/13 - DMF (solid), S/13 - GMF
(dotted), S/17 - DMF (dashed), and S/17 - GMF (dashed-dotted),
respectively, where in the latter two samples event number 22 has been
excluded.}
\end{figure}

\section{Conclusions}

In this paper, we have shown that the most straightforward
interpretation of the existing microlensing observations of stars
within the LMC, assuming the majority of candidate events to be due to
compact lenses within the MW halo, may actually be in conflict with
the observational data itself.  The MACHO collaboration uses a
likelihood analysis to derive two "best-fit" parameters from their
microlensing observations: the typical MACHO mass and MACHO halo
fraction.  A naive comparison of the theoretical event rate
distribution produced by these "best-fits" assuming a standard halo
model, with an ad-hoc Gaussian event rate distribution indicates that
the spread in the durations of the observed events seems smaller than
would be expected for a standard halo of compact dark matter. In order
to quantify this suspicion we have formulated a statistic, the $Width$
(essentially the spread in the observed event durations relative to
the average event duration), that is virtually independent of the
MACHO best-fit parameter values.

We have carried out Monte Carlo simulations to calculate the expected
distribution of the $Width$, under the assumption of a standard halo
MACHO population, utilizing two different background subtraction
methods (fixed number and Poissonian), for both detection criteria,
for delta function and finite width MACHO mass functions and excluded
various sets of events suspected of being due to background populations. The
probability of observing a value of the $Width$ as small as 
observed falls, in {\it all} cases, below $10\%$, with values often
below $5\%$. In other words the
underlying model, in terms of the expected background event rate and a
standard halo MACHO population, is excluded at between the $90\%$ and
$95\%$ confidence level. 

The discrepancy between the observational data and its halo dark
matter interpretation may be alleviated if the number of events which
are due to non-halo populations is significantly larger than expected,
at the expense of correspondingly reducing the MACHO halo
fraction. Such a reasoning also implicitly assumes the non-halo
population to produce microlensing events with event durations being
by chance very similar to those of the halo population. Possibly more
natural would be an explanation in terms of "all background", though
the nature of this background, and the reason for it producing such a
narrow distribution in durations, remains to be established.
Alternatively, the discrepancy may indicate that the standard halo
model is a poor approximation to the actual MACHO distribution (Widrow
\& Dubinski~\cite{sim}). For instance, there could be a clump with
small intrinsic velocity dispersion crossing the line of sight to the
LMC, which, nevertheless, would be required to be surprisingly
massive.  Another possible resolution to the problem could be that the
statistical correction to the event durations, due to blending, leads
to a reduction in the spread of the durations i.e. that the underlying
durations of some observed events are significantly longer than their
estimated values. If this were the case, however, then conclusions
based on these estimated timescales, not only the analysis in this
paper, but also the derivation of exclusion limits on the halo
fraction in massive black holes (Alcock et al.~\cite{long}) due to the
purported absence of long-duration events, would seem unreliable.

Further clarification of the nature of the lenses may hopefully
result from an enlarged sample of microlensing candidates, as expected
from the upcoming analysis of the MACHO 8 year data and EROS 5 year.
Evidently a longer survey duration will increase the number of
candidate events, reducing the ambiguity associated with Poisson
statistics, as well as increasing the efficiency for detecting longer
duration events, thus making the width of the distribution a
significant discriminator between models.  Nevertheless, a signifcant
reduction of Poisson ambiguity may only result by an increase of the
number of events by an order of magnitude, such as anticipated in a
next generation SUPERMACHO survey (Stubbs~\cite{stubbs}).  This later
survey is expected to also reduce uncertainties in event duration
corrections due to blending as it is planned to employ the
technique of difference image analysis.


\begin{acknowledgements}
We are grateful to David Bennett, Alain Milsztajn and
Esteban Roulet for useful comments. AMG was supported by PPARC and the
Swedish Research Council, and acknowledges the use of the starlink
computing facilities at QMW.     
\end{acknowledgements}

\end{document}